\documentclass{webofc}
\usepackage[varg]{txfonts}   
\usepackage{hyperref}
\usepackage{url}
\usepackage{subcaption}
\hypersetup{colorlinks=true,citecolor=blue,urlcolor=blue,linkcolor=blue}
\begin{document}
\title{Measurements of Light Nuclei (d, t, $^3$He)--$\Lambda$ Correlations in Au+Au Collisions at $\sqrt{s_{NN}}=3$ GeV from STAR}
\author{\firstname{Xialei} \lastname{Jiang}\inst{1,2}\fnsep\thanks{\email{xialeij@mails.ccnu.edu.cn}}(For the STAR Collaboration)
}

\institute{Key Laboratory of Quark \& Lepton Physics (MOE) and Institute of Particle Physics, Central China Normal University, Wuhan 430079, China
\and
           GSI Helmholtzzentrum für Schwerionenforschung, 64291, Darmstadt, Germany
          }

\abstract{Heavy-ion collisions offer a unique way to study hyperon–nucleon ($Y$–$N$) interactions through two-particle momentum correlations, which reveal the source’s space-time structure and the effects of the final state interactions. Correlations between light nuclei (d, t, $^{3}$He) and $\Lambda$ provide insight into hypernuclei structure, binding energies, and many-body interactions that might be relevant to the inner structure of neutron stars. This work presents the first measurements of d–$\Lambda$, t–$\Lambda$, and $^{3}$He–$\Lambda$ correlations from $\sqrt{s_{_{\rm NN}}} = 3$ GeV Au+Au collisions collected in 2021 at STAR. Using the Lednicky–Lyuboshitz model, we extract source sizes and interaction parameters, shedding light on hyperon interactions and light hypernuclei structure.
}
\maketitle
\section{Introduction}
\label{intro}

Femtoscopy, inspired by the Hanbury Brown–Twiss (HBT) effect \cite{HBT1956}, is a technique that probes the space-time structure of the particle-emitting source and final state interactions (FSI) through momentum correlations between particle pairs. The theoretical correlation function is expressed as:
\begin{equation}
C_{th}(k^*) = \int d^3r^* S(\vec{r}^*) |\Psi(\vec{r}^*, \vec{k}^*)|^2
\label{therotical formalism}
\end{equation}
where $S({r}^*)$ is the source function and $\Psi({r}^*, {k}^*)$ is the pair wave function.  
Experimentally, the signal distribution $A(k^*)$ is constructed by pairing particles within the same event, while the reference distribution $B(k^*)$ is formed using mixed events to remove physical correlations. The measured correlation function is then:
\begin{equation}
C_{exp}(k^*) = N \cdot \frac{A(k^*)}{B(k^*)}
\label{experimental formalism}
\end{equation}
with N as a normalization constant. The Lednicky–Lyuboshitz model uses Effective Range Expansion (ERE) to characterize the scattering process with only two parameters: the scattering length $f_{0}$ and the effective range $d_{0}$\cite{lednicky1981}. 

\section{Correlation Function and Discussion}
\label{sec-1}
The STAR collaboration has collected approximately 2 billion Au+Au collision events at $\sqrt{s_{NN}}=3$ GeV in fixed-target mode, taken as part of the Beam Energy Scan Phase II. This high-statistics dataset enables the measurement of t–$\Lambda$ and $^3$He–$\Lambda$ correlation functions. The Time Projection Chamber (TPC) with iTPC upgrade and Time-Of-Flight (bTOF) detectors, are used to identify particles. $\Lambda$ hyperon are reconstructed using the Kalman Filter Particle (KFP) method \cite{ju2023}.
\subsection{d--$\Lambda$ Correlation}
\label{sec-2}
\begin{figure}[h]
    \centering
    \includegraphics[width=0.7\linewidth, height=4.9cm, keepaspectratio]{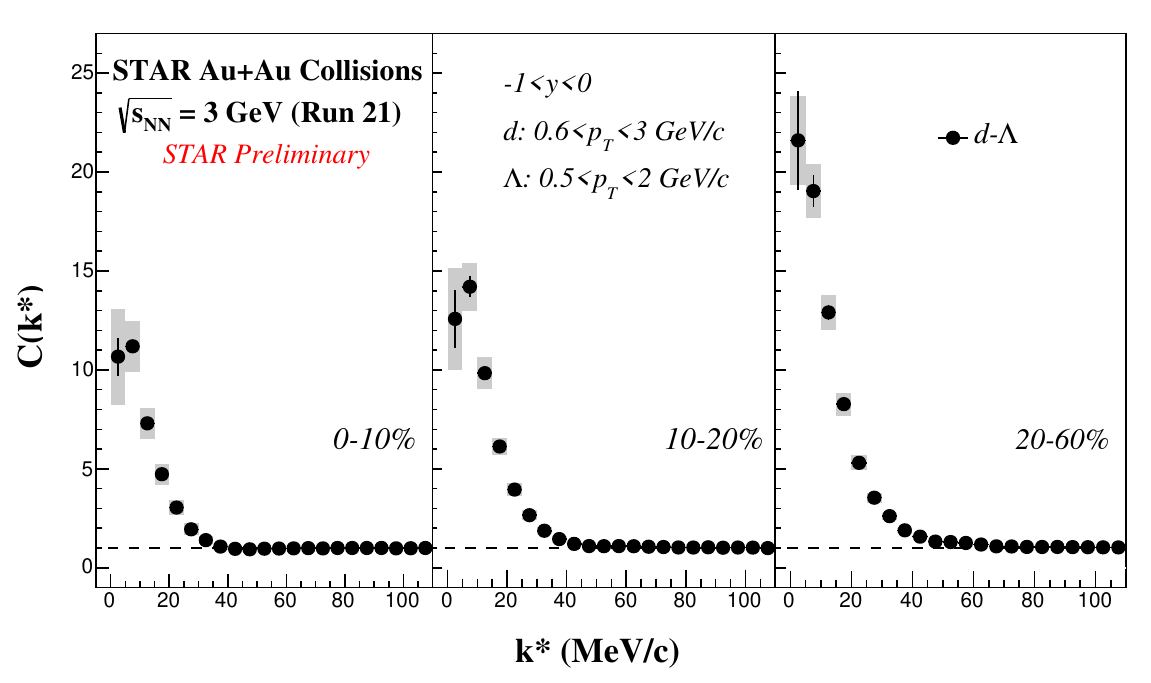}
    \caption{Measured d-$\Lambda$ correlation functions in 0-10\% (left), 10-20\% (middle) and 20-60\% (right) centrality. The black vertical bars and boxes represent the statistical and systematic uncertainties, respectively.}
    \label{fig:dLambda cf}
\end{figure}

\begin{figure}[h]
  \centering
  \begin{subfigure}[t]{0.48\textwidth}
    \centering
    \includegraphics[height=5.4cm, width=1.5\linewidth, keepaspectratio]{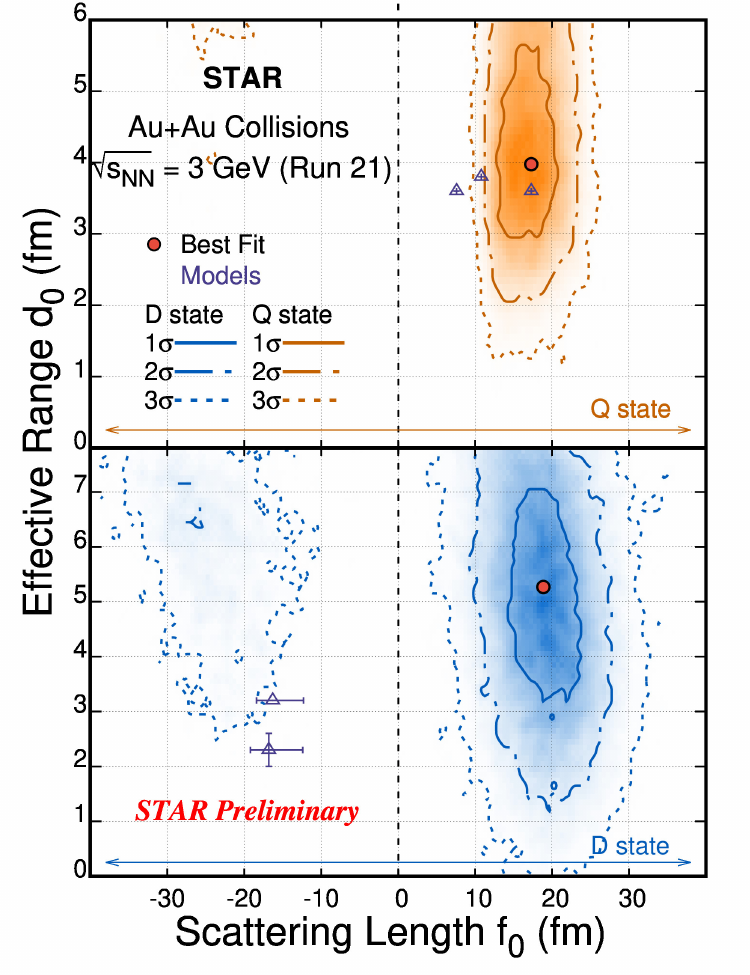}
    \caption{Unconstrained LL fit contour}
    \label{fig:unconstrained LL fit}
  \end{subfigure}
  \hfill
  \begin{subfigure}[t]{0.48\textwidth}
    \centering
    \vspace{-4.2cm}
    \includegraphics[height=5cm, width=\linewidth, keepaspectratio]{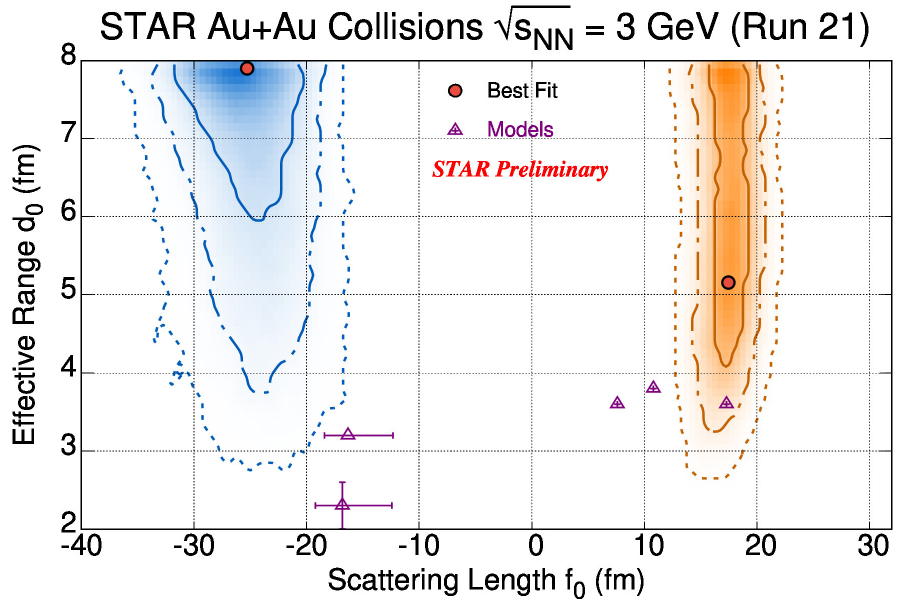}
    \caption{Constrained LL fit contour}
    \label{fig:constrained LL fit}
  \end{subfigure}

  \caption{Contours of the probability distributions for the extracted final-state interaction (FSI) parameters — scattering length ($f_{0}$) and effective range ($d_{0}$) - for the d-$\Lambda$ system are shown for both the doublet (D, blue) and quartet (Q, orange) spin states. The purple triangles represent predictions from various theoretical models.}
  \label{fig:dLambda LL fit contour}
\end{figure}

\begin{figure}[h]
    \centering
    \includegraphics[width=0.7\linewidth, height=4.5cm, keepaspectratio]{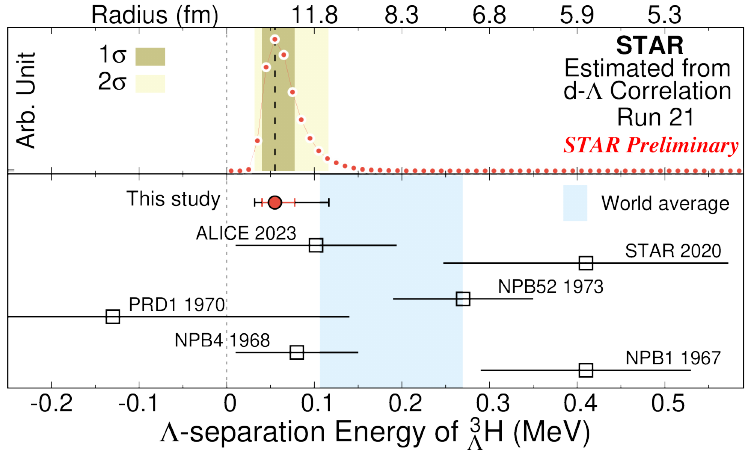}
    \caption{$\Lambda$ separation energy of $^{3}_{\Lambda}$H ($B_{\Lambda}$). The upper panel displays the probability distribution of the $\Lambda$ separation energy $B_{\Lambda}$. The red point in the lower panel indicates the value of $B_{\Lambda}$ measured in this analysis, with a 95\% confidence level. Previous global measurements are also shown for comparison, and their weighted average, $0.19 \pm 0.06$ MeV, is represented by the blue band.}
    \label{fig:H3L BE}
\end{figure}

Figure~\ref{fig:dLambda cf} shows the d–$\Lambda$ correlation function with systematic uncertainties. A pronounced enhancement at low relative momentum $k^*$ reflects a strong attractive interaction between deuterons and $\Lambda$ particles, with the correlation strength varying across collision centralities. It is important to note that the d–$\Lambda$ correlation also contains contributions from the three-body decay of $^3_\Lambda$H ($^3_\Lambda$H → p$\pi^-$d). Due to the similarity in decay signatures and limited experimental resolution, these contributions cannot be distinguished from genuine $\Lambda$ decays. Based on STAR measurements and simulations, this contamination accounts for 4–11\% of the d–$\Lambda$ pairs at $k^* < 100$ MeV/c and has been properly subtracted from the analysis.
We fit the data using the Lednicky–Lyuboshitz (L–L) model, considering the doublet and quartet spin states with weights of 1/3 and 2/3, respectively.In the “unconstrained L–L fit,” the scattering length $f_0$ is allowed to vary freely over positive and negative values. Bayesian fitting results (Fig.~\ref{fig:unconstrained LL fit}) show that $f_0$ is more likely positive, around 18 fm, for both spin states. However, the doublet state also shows some probability for negative $f_0$. Assuming the existence of the $^{3}_{\Lambda}$H bound state, the doublet $f_0$ should be negative. Therefore, we constrain its range to [–50 fm, 0 fm] (Fig.~\ref{fig:constrained LL fit}). Due to the limited precision in extracting the effective range $d_0$, only the scattering length $f_0$ for each spin state is reliably determined.
Using the effective range expansion formalism, the binding energy of $^{3}_{\Lambda}$H is related to the fitted parameters $f_0$ and $d_0$ in the doublet channel. The upper panel of Fig.~\ref{fig:H3L BE} illustrates the probability distribution of the binding energy alongside the corresponding inferred radius, while the lower panel compares our results with previously reported global measurements. Our analysis yields a binding energy of $0.06_{-0.02}^{+0.06}$ MeV/c² and an associated radius of $16_{-5}^{+5}$ fm, consistent with the world average and characterized by a notably small uncertainty(95\% CL). \cite{Chen:2023mel}.
\subsection{t--$\Lambda$ and $^{3}$He--$\Lambda$ Correlation}
\begin{figure}[h]
    \centering
    \includegraphics[width=0.9\linewidth, height=4.5cm, keepaspectratio]{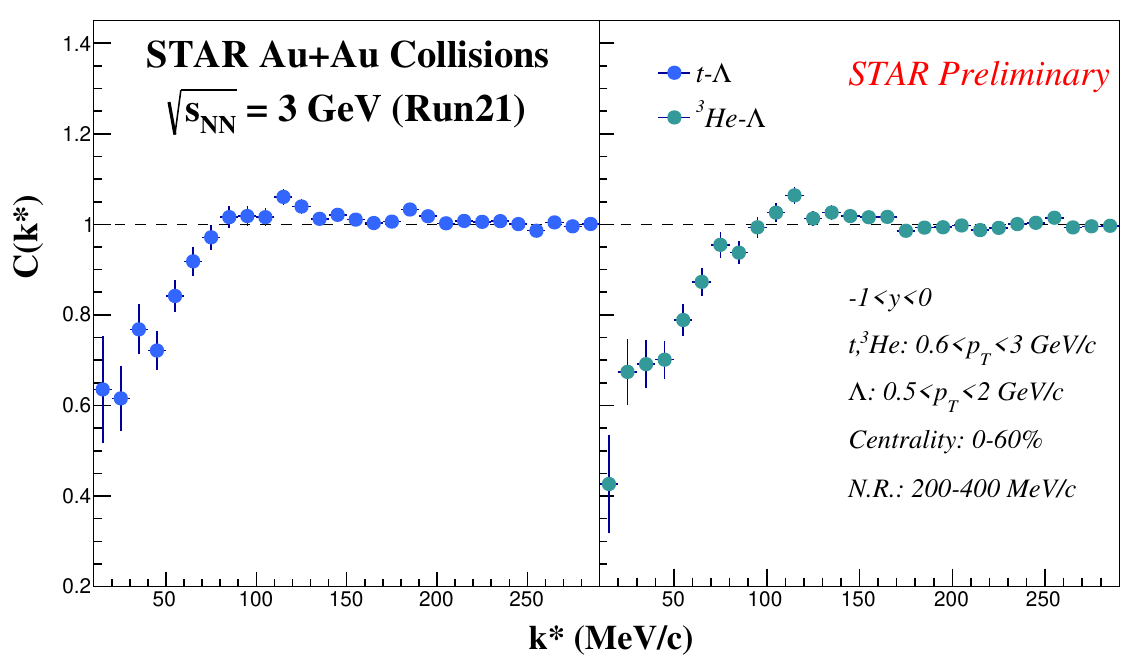}
    \caption{Measured t-$\Lambda$ (left) and $^3$He-$\Lambda$ (right) correlation functions for 0–60\% centrality. Black vertical bars indicate statistical uncertainties. Systematic uncertainties are not included in the figure.}
    \label{fig:the3Lambda correlation function}
\end{figure}

We also present the correlation function results for t–$\Lambda$ and $^{3}$He–$\Lambda$. Similar to how $^{3}_{\Lambda}$H contamination affects the d–$\Lambda$ correlation function, daughter $p$ and $\pi^{-}$ from $^{4}_{\Lambda}$H and $^{4}_{\Lambda}$He decays are misidentified as fake $\Lambda$ particles, thereby contaminating the t–$\Lambda$ and $^{3}$He–$\Lambda$ correlation functions. To investigate this effect, we divided the invariant mass distribution of $\Lambda$ candidates within a $\pm$2$\sigma$ range into two regions and studied their impact. For the correlation functions presented here, we use only $\Lambda$ candidates in the invariant mass window 1.11589–1.11906 GeV/$c^{2}$, which minimizes the contamination from hypernuclei decays. As illustrated in Fig.~\ref{fig:the3Lambda correlation function}, the t–$\Lambda$ and $^{3}$He–$\Lambda$ correlation functions exhibit similar structural characteristics. Notably, an enhanced structure appears near $k^{*} = 120$ MeV. It remains to be determined whether this feature arises from particle decay or from final state interactions. Further detailed analysis is required to clarify these contributions and to deepen our understanding of the underlying physics.
\section{Summary and Outlook}
This proceeding reports the latest result of the d–$\Lambda$ correlation measurement at $\sqrt{s_{NN}} = 3$ GeV from the STAR experiment. Based on the correlation analysis, the binding energy of $^3_{\Lambda}$H is estimated to be $0.06_{-0.02}^{+0.06}$ MeV/c², with a corresponding radius of approximately $16_{-5}^{+5}$ fm. The result is consistent with the world average and features a notably smaller uncertainty. In addition, this work presents the first measurements of the t–$\Lambda$ and $^{3}$He–$\Lambda$ correlation functions. In the future, relevant scattering parameters will be extracted to estimate the binding energies of $^4_{\Lambda}$H and $^4_{\Lambda}$He.
\section{Acknowledgement}
This work was supported by National Key Research and Development Program of China (NO. 2022YFA1604900), National Natural Science Foundation of China (NO. 12525509 and 12447102).

\end{document}